\documentclass[ag]{copernicus}

\usepackage{color}

\begin{document}

\title{{Downward transport of electromagnetic radiation by electron holes?}}

\author[1,2]{R. A. Treumann}
\author[3]{W. Baumjohann}
\author[2]{J. LaBelle}
\author[4]{R. Pottelette}

\affil[1]{Department of Geophysics and Environmental Sciences, Munich University, Munich, Germany}
\affil[2]{Department of Physics and Astronomy, Dartmouth College, Hanover NH 03755, USA}
%\affil[3]{International Space Science Institute, Bern, Switzerland}
\affil[3]{Space Research Institute, Austrian Academy of Sciences, Graz, Austria}
\affil[4]{LPP, Saint Maur des Foss\'es, France}

\runningtitle{Downward radiation transport}

\runningauthor{R. A. Treumann, W. Baumjohann, J. LaBelle \& R. Pottelette}

\correspondence{R. A.Treumann\\ (rudolf.treumann@geophysik.uni-muenchen.de)}

\received{ }
%\pubdiscuss{ } %% only important for two-stage journals
\revised{ }
\accepted{ }
\published{ }

%% These dates will be inserted by the Publication Production Office during the typesetting process.

\firstpage{1}

\maketitle

\begin{abstract}
An attractive mechanism for radiation transport by electron holes from the magnetospheric auroral cavity source region down to the ionosphere and possibly even down to the atmosphere is examined. Because of the excitation and propagation properties of the X mode, this mechanism turns out to be highly improbable for the usual resonant excitation of radiation of frequency just below the local electron cyclotron frequency, $\omega\lesssim\omega_{ce}$. It could work only, if the auroral ionosphere would be locally perforated being of sufficiently low density for allowing electron holes  riding down to low altitudes on the auroral electron beam. If resonant excitation of higher electron cyclotron harmonics, $\omega\sim \ell\omega_{ce}, ~\ell= 2,3,\dots$, becomes possible, a still unexplored mechanism, then radiation excited inside the hole could be transported to lower altitudes than generated. If this happens, radiation would provide another mechanism of coupling between the magnetospheric plasma and the atmosphere.

\end{abstract}

\introduction

This Letter shows that the otherwise very attractive idea of transport of narrow band auroral kilometric radiation generated inside electron holes \citep[the generation mechanism has been elucidated in][]{treumann2011a,treumann2011b} in the auroral cavity  down to the ionosphere and possibly even further down to the neutral atmosphere is (probably) quite unrealistic. 

Such a radiation transport might occur just under extremely favorable circumstances prevalent in the ionosphere when the ionosphere becomes perforated. Unfortunately, under any other or normal conditions, however, electromagnetic radiation transport by holes, as can naively be guessed, will probably not take place.  We do not explore the transport mechanism in detail, for a detailed calculation requires extended access to the space-time distribution of the electron densities and temperatures at altitudes reaching from the atmosphere up to the magnetosphere, all longitudes and latitudes in the interval from subauroral to polar latitudes. This effort is outside the scope of the present communication. However, we feel that the mere idea is interesting and tempting enough for excluding it as a possible radiative coupling mechanism between the atmosphere and magnetosphere from upside down, at least at frequencies below $\omega_{ce}$.

Coupling between the atmospheric and plasma environments of Earth has recently become of increasing interest. There is no doubt that such coupling really exists. It is, however, believed that it is mainly upward and of basically mechanical and thermal nature. The various known coupling processes have recently been reviewed \citep{bosinger2012} to some detail. Some surprising effects of Earth's geomagnetic field on the atmospheric density distribution have been discovered by the CHAMP spacecraft \citep{luhr2012}.  Otherwise the effect of the magnetospheric plasma on the atmosphere is probably restricted mainly to auroral and polar latitudes and depends heavily on the magnetospheric particle distribution and its space-time and energy-time variability. 

On the other hand, from a magnetospheric point of view,  some violent processes seem not to have any known or remarkable effect in the ionosphere and atmosphere. Such processes relate to the generation of auroral kilometric radiation which is known to contain several per cent of the total energy content of a magnetospheric substorm, which in radiation is a huge percentage. Though this number is still energetically negligible in any large-scale atmospheric or ionospheric process, the question arises whether it could be transported at all down to the ionosphere and atmosphere where it would be detected as banded free-space radiation. Interestingly, \citet{labelle2011b} recently reported the probable observation of auroral kilometric radiation even at the ground from Antarctica. In the upper ionosphere banded (so-called MF burst) radiation has also been observed though it is not yet finally decided whether it is electromagnetic or electrostatic \citep{labelle2011a}.\footnote{For plasma wave observations see \citet{labelle1988,labelle2002}.} Its generation mechanism remains still unclear as also its relation to substorms. Here, we take these just as examples while not claiming that they could be explained by the mechanism we are going to check below. 

\section{Electron hole generated radiation} 
In two recent papers \citep{treumann2011a,treumann2011b} electron holes in the auroral upward and downward \citep[for observational evidence see, e.g.,][]{carlson1998,ergun1998a,ergun1998b,ergun1998c,ergun2002} current have been identified as being potentially responsible for the generation of the well-observed fine structure in the auroral kilometric radiation. This radiation is generated by the electromagnetic electron-cyclotron maser \citep{wulee1979} mechanism \citep[for a review cf., e.g.,][]{treumann2006}. 
In the former papers it was recognised that finestructure of such radiation can preferably be generated in the steep density gradient between the interior of electron (and possibly though to a lesser extent also ion) holes in the auroral plasma and the ambient plasma. Electron holes, as their name tells, are  deficient of electrons. Thus the short spatial scale density gradient at the hole boundary is steep and, in the presence of fast electrons, contributes to the electron cyclotron maser. Generated on the inside of the hole, the radiation becomes trapped by the hole thus staying in long-time contact with the gradient. It can potentially grow to large amplitudes until achieving substantial intensity. Radiation generated by electron holes is about \emph{precisely perpendicular} thereby favouring trapping, excitation and amplification. In principle, resonance at higher cyclotron harmonics would also be possible. Such radiation at auroral altitudes had frequency high enough to escape from the hole but in the denser external plasma at lower altitudes could as well be trapped.

The fact that radiation is trapped by the hole opens up the possibility for its transport along the magnetic field. We should note that this transport may not be strictly parallel to the field, because holes amplify the ambient field, possess finite magnetic moments, and thus may experience electric field drifts which shift them away from the original field line \citep{treumann2012}. This effect is, however, secondary  in the present context. It would contribute to latitudinal displacement of the radiation source from the auroral region preferentially to higher (polar) latitudes. 

Electron holes ride on the fast electron component along the magnetic field though at somewhat lesser velocity $V_h<v_b$, with $v_b$ the electron beam speed \citep{treumann2008}. In the downward current region the beam and holes flow upward from the ionosphere into the magnetosphere. They have been made responsible for most of the observed finestructure in the auroral kilometric radiation \citep{treumann2011b}. 

In the upward current region, being generated by the fast downward directed auroral electron beam, holes necessarily move downward together with the auroral electrons.  Such holes have been identified to exist in the upward current region \citep{pottelette2005}.  They are in dynamical pressure equilibrium between external plasma and the trapped electron component. They are the candidates which interest us here. If they are capable of making it down to the deeper layers of the ionosphere or even the atmosphere they would be the ideal candidates for transport of the trapped auroral electromagnetic radiation produced by and in them down to low latitudes where no radiation can be generated\footnote{With the possible exception of lightning strokes.} because either of (a) lack of a relevant mechanism (in the atmosphere), (b)  of the impossibility of propagation of any radiation in the (ionospheric) plasma. 

\begin{figure}[t!]
\centerline{\includegraphics[width=0.5\textwidth,clip=]{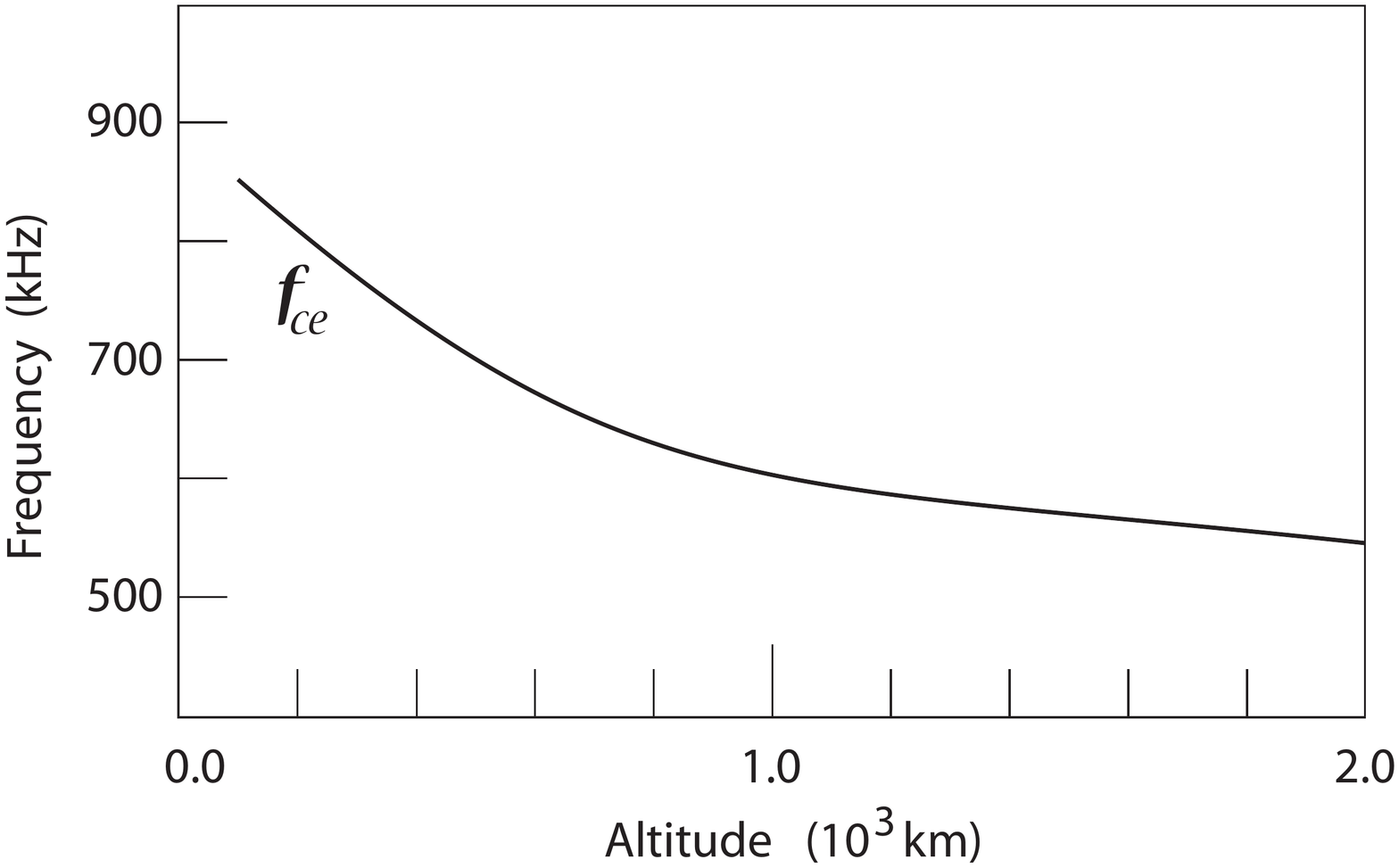}}\vspace{-1cm}
\caption[ ]
{\footnotesize {Altitude variation of electron cyclotron frequency $f_{ce}=\omega_{ce}/2\pi$.}}
\label{fig-fce-vers-h}
\end{figure}

\section{Downward transport}
In order to transport the trapped electromagnetic radiation down into the ionosphere, an electron hole born in the upward current region must survive for the time it needs to traverse the roughly 2000 km to 4000 km difference in altitude. 

Formation of electron holes is a fast process. Numerical simulations performed under idealised non-magnetic plasma conditions \citep{newman2002,ergun2002} valid along the ambient magnetic field suggest that their lifetime is of the order of $\tau\omega_e\sim10^4$ or longer. Plasma frequencies in the upward-current hole-generation region are low. The ambient field aligned electric potential evaporates the electron background. The plasma density is thus solely due to the fast auroral electrons, yielding plasma frequencies in the range of few kHz up to $\lesssim15$ kHz \citep{pottelette2005}, and implying idealised hole  lifetimes of the order of around $\tau\lesssim 1$ s. 

Riding on the downward auroral beam the holes move slightly slower than the beam. Assuming about $\lesssim 10$ keV auroral electrons, which is frequently observed, corresponds to downward electron velocities of $\lesssim 0.1 c$ or travel times of $\sim30$ ms per $10^3$ km along the magnetic field. The upper atmospheric absorption altitude of $\sim10$ keV electrons  is located around $\sim90$ km, just below the ionospheric E-layer, where the plasma density of the D-layer is (practically) negligible compared with the density of neutrals and collisional absorption dominates. Starting at around $\sim3000$ km altitude, the electrons reach this location within roughly $\sim90$ ms. A slightly slower surviving electron hole would thus need, say, $\Delta t\gtrsim 100$ ms to penetrate this deep into the atmosphere, a time which is well within its nominal life time $\Delta t<\tau$.

Theoretically, an electron hole could thus reach atmospheric altitudes if it would ride down the electron beam up to the absorption height. 

This, however, seems improbable for several reasons, the main of which is that, at F-layer altitudes, the electron beam becomes a minor distortion of the dense ionospheric background plasma losing its property of being the dominant electron component which it had in the auroral cavity, the presumable source region of auroral kilometric radiation. 

By the property of the electron hole to adjust to the bulk speed of the main plasma component \citep{treumann2008} the downward hole velocity should readily slow down when the auroral electron beam enters the upper F-layer from above leaving the auroral cavity. This causes the hole to stay behind the beam and disperse. If it would indeed manage to pass the ionosphere and enter the neutral atmosphere then, after destruction of the hole, it could release the trapped radiation as a free electromagnetic wave at frequency close to the last excited radiation below the local $\omega_{ce}$, somewhere in the several hundred kHz range, propagating on the X mode in the upper atmosphere. Latitudinally, its source would be located below the aurora.

A more realistic scenario should take into account the propagation properties of the electron hole, radiation production and propagation.

\section{Improbability of radiation transport}
Emitted, trapped, amplified, and reabsorbed radiation in the X mode can propagate inside the hole only on the X mode branch. Radiation is emitted at frequency $\omega_{ce}>\omega>\omega_X$ below the electron cyclotron frequency $\omega_{ce}$ and above the X-mode cut-off $\omega_X$.

As long as the electron beam \emph{inside} the hole is the dominant dilute plasma component one has $\gamma>1$ about constant, and, for $\omega_e<\omega_{ce}$,  the X mode branch cut-off $\omega_X\sim \omega_{ce}/\gamma$, with $\gamma$ the relativistic factor, drops some distance in frequency below the local electron cyclotron frequency $\omega_{ce}$ \citep[cf., e.g.,][]{pritchett1984a,pritchett1984b,pritchett1984c,pritchett1986}, allowing the radiation to propagate. During downward displacement of the hole, however, the electron cyclotron frequency increases steeply (see Fig. \ref{fig-fce-vers-h}). Thus, radiation at the low auroral kilometric radiation frequencies ceases to propagate in the X mode once the frequency drops below the X-mode cut-off. 

If no low-energy plasma leaks in on the displacement time scale into the hole, the condition for the maser cyclotron instability will still be satisfied, causing excitation of ever higher electromagnetic frequencies inside the hole at the steep density gradient, with the low frequency part of the spectrum gradually dropping below cut-off and disappearing. 

On the other hand, for low internal hole density one has $\omega_{ce}\sim\omega_{uh}$, the internal upper hybrid frequency, and the low-frequency X-mode might tunnel and jump over to the short-wavelength Z-mode branch where it propagates as long as its frequency stays above the lower Z-mode cut-off $\omega_Z>\gamma\omega_e^2/\omega_{ce}$, where $\omega_e$ is the plasma frequency based on the internal hole density.

On the other hand, at the altitude where cold plasma from the outside leaks in, the hole will become destroyed and the radiation absorbed. Since, inside the hole, the co-moving electron cyclotron maser primarily does not excite frequencies above the electron cyclotron frequency for purely perpendicular propagation (which is the case in the hole frame). No co-moving Z-mode radiation will be released. This case is thus not of any vital interest. 

The most serious obstacle of downward radiation transport is, however, provided by the propagation properties of the electron hole itself.  Holes do not propagate at a constant velocity for all their life times. The bulk velocity of the hole in the external space (i.e. the ionosphere) is determined by the presence of the adjacent plasma external to the hole. As long as this adjacent external component is dominated by the electron horseshoe beam of the auroral electrons with practically no electron background, the electron hole tries to adjust its speed to the beam speed. This is reflected in numerical simulations of electron holes \citep{newman2001,newman2002}. However, when the holes enter the dense though still collisionless ionosphere external to the hole, the density of which strongly exceeds the density of the residual dilute high energy auroral electron beam, the bulk speed of the hole will drastically drop. The hole will retard coming to rest. The electron cyclotron frequency at the altitude where this happens determines the final frequency which the electron cyclotron maser generates, a frequency somewhere in the range well below $< 1$ MHz.  When the hole decays, this radiation will be released but can propagate at best if its frequency and wave number satisfy the local conditions for Z-mode radiation.

\conclusions
From the previous we are thus left with the case when the electron hole can really make it down to the atmosphere, crossing the ionosphere, i.e. when accidentally the ionosphere becomes perforated at auroral latitudes with local densities sufficiently low for the external plasma not to retard the hole on time scales shorter than downward propagation. Only in these rare cases with the hole riding down on the $\sim10$ keV auroral electron beam, it would be able to transport the internally generated electromagnetic radiation down to the lower ionosphere or even the atmosphere. The frequency of such radiation would always be close to the local electron cyclotron frequency as given in Figure \ref{fig-fce-vers-h}. Without extended numerical calculations using realistic auroral electron density profiles under various conditions it is not possible to conclude whether or not such a transport of radiation by electron holes can indeed take place.  In such presumably very rare cases of ionospheric perforation escaping radiation at D-region altitudes would occur as narrow banded free space electromagnetic radiation and might explain observations of the kind as reported by \citet{labelle2011b}. The probability for this to happen is very low. 

On the other hand, the case when radiation is resonantly generated inside the hole at higher electron cyclotron harmonics $\omega\sim\ell\omega_{ce},~\ell=2,3,\dots$ and becomes trapped at sufficiently low density where the external plasma density is high enough to inhibit escape from trapping, radiation may well become transported down. Such kind of radiation generation has so far not yet been considered.

However, in completely different situations, when electron holes are excited in lightning discharges, generation and transport of electromagnetic radiation near the local cyclotron frequency across the lightning channel should cause escape to the atmospheric environment.

\begin{acknowledgements}
This research was part of an occasional Visiting Scientist Programme in 2006/2007 at ISSI, Bern. RT thankfully recognises the assistance of the ISSI librarians, Andrea Fischer and Irmela Schweizer. He appreciates the encouragement of Andr\'e Balogh, Director at ISSI.
\end{acknowledgements}

\end{document}